\documentstyle[sprocl,cite,epsfig]{article}

\def\bbox#1{\mbox{\boldmath $#1$}}

\begin{document}

\title{
PION PRODUCTION MODEL - CONNECTION BETWEEN DYNAMICS AND QUARK MODELS}

\author{T.-S. H. LEE}

\address{Physics Division, Argonne National Laboratory, Argonne,
Illinois 60439,
U.S.A. \\ E-mail: lee@anlphy.phy.anl.gov}

\author{T. Sato}

\address{Department of Physics, Osaka University, Osaka 560-0043, Japan \\
E-mail: tsato@phys.sci.osaka-u.ac.jp}


\maketitle\abstracts{We discuss the
 difficulties in testing the hadron models by using
the $N^*$ parameters extracted
from the empirical amplitude analyses of the
$\pi N$ and $\gamma N$ reaction data.
As an alternative or perhaps a more advantageous approach, 
we present a Hamiltonian formulation
that can relate the pion production dynamics and
the constituent quark models of $N^*$ structure. 
The application of the approach in investigating
the $\Delta$ and $N^*(S_{11})$ excitations is reviewed. 
It is found that while the $\Delta$ excitation can be described
satisfactory, 
the $\pi N$ scattering in $S_{11}$ channel can not be
described by the constituent quark models based on either the 
one-gluon-exchange or one-meson-exchange mechanisms.
A phenomenological quark-quark potential has been constructed to
reproduce the $S_{11}$ amplitude.} 

One of the challenging theoretical problems is to understand the
hadron structure within Quantum Chromodynamics(QCD). There exist two different approaches.
The most fundamental one is the Lattice QCD calculation. The second
one is to develop various QCD-inspired models. While a lot of
progresses have been made in this direction, 
only very limited attention has been paid to the need of
developing appropriate reaction
theories for
testing their predictions by using the data of $\pi N$ and $\gamma N$
reactions.
In the past few years, we have addressed this question concerning
the constituent quark models. Here we would like to review the
progresses we have made\cite{satolee1,satolee2,yoshimoto} and
discuss future directions. The
work for $N^*(S_{11})$ involves a collaboration with T. Yoshimoto
and M. Arima.
 
We were motivated by the following observation.
It is  common to compare the masses and decay widths predicted
by the constituent quark models(or any existing hadron models)with
the data listed by Particle Data Group(PDG).  
All calculations of decay widths have been 
done\cite{copley,foster,koni,li,iach,caps2} 
perturbatively. 
It has been found that such a perturbative calculation can at best describe 
the general 
qualitative 
trend of the data, but not the quantitative
details. It is important to note here that the PDG's values
are extracted from the empirical  $\pi N\rightarrow \pi N$ and
$\gamma N \rightarrow \pi N$ amplitudes which
contain both resonant and non-resonant components. In most partial waves,
the non-resonant mechanisms are important
; one can see this from the fact
that  most of the resonances identified by PDG are in fact
not visible in $\pi N$ and $\gamma N$ cross section data.
By the unitarity condition, therefore the extracted resonance parameters
$inherently$ contain non-resonant contributions.  
Clearly, except in 
a region where the non-resonant contributions are
negligibly small, the comparison of the 
PDG values (or values from other amplitude analyses) with the 
decay widths calculated perturbatively from the constituent quark models
could be very misleading. In particular, a perturbative calculation
of decay widths is obviously not valid for 
cases in which
two nearby resonances in the same partial wave 
can couple with each other through their coupling with
the meson-nucleon continuum.
Similar precautions must also be taken 
in comparing the predicted masses with the PDG values. 

To have a more direct 
test of constituent quark models, it is necessary to develop
a nonperturbative approach that 
takes account of
the unitarity
condition and can relate the $\pi N$ 
and $\gamma N$ reactions directly to
the predicted internal quark wave functions of baryons. 
We have achieved this by developing an approach within
the Hamiltonian formulation.
In the following, we will first describe the major steps and discuss how
our previous work\cite{satolee1} on the $\Delta$ excitation
can be interpreted within this framework. We then
discuss our findings from an investigation of the excitation of
the $N^*(S_{11})$ resonances in $\pi N$ scattering.

We start with the usual constituent quark model defined by the
following Hamiltonian
\begin{equation}
  h_B = K + V_{\rm conf} + V_{qq}.   
\end{equation}
 where $K$ is the kinetic energy, 
$V_{\rm conf} = \sum_{i<j} \alpha_c r_{ij}$ is the usual linear 
confinement potential.
For the residual $qq$-interaction $V_{qq}$ in Eq.(1), 
we consider both the usual one-gluon-exchange(OGE) 
model\cite{isgur,caps1,capstick} and
the recently developed\cite{gloz1,gloz2} one-meson-exchange(OME) 
model(the weak $\eta$-exchange is suppressed here, but it 
was included in our investigation\cite{yoshimoto}).
Explicitly, we have 
\begin{eqnarray}
  V^{OGE}_{qq} &=& \sum_{i < j}<\lambda_i\cdot\lambda_j >
 \left[\bbox{\sigma}_i \cdot \bbox{\sigma}_j
  V^g_{\sigma}(r_{ij})
  + S_{ij} V^g_{T}(r_{ij}) \right],
\end{eqnarray}
\begin{eqnarray}
  \label{potential}
  V^{OME}_{qq} &=& \sum_{i < j} 
  \bbox{\tau}_i \cdot \bbox{\tau}_j
\left[\bbox{\sigma}_i \cdot \bbox{\sigma}_j
  V^\pi_{\sigma}(r_{ij})
  + S_{ij} V^\pi_{T}(r_{ij}) \right],
\end{eqnarray}
where 
the color SU(3) factor is $<\lambda_i\cdot\lambda_j> = -\frac{8}{3}$,
 $\bbox{\sigma}_i$ and $\bbox{\tau}_i$ are respectively the spin and 
isospin operators,
and $ S_{ij}$ is the usual tensor operator. 
 The radial parts of the potentials
in Eqs.(2)-(3) are given in Ref.\cite{yoshimoto}.
We only note here that they are 
regularized by form
factors
  $F^{\alpha}(\bbox{q}) = \Lambda^2_{\alpha}/(\Lambda^2_{\alpha}
 + \bbox{q}^2)$. 
This is consistent with the notion that the constituent quarks are not point
particles within an effective theory. This regularization
of the $qq$-potential is essential in obtaining convergent solutions
for the bound state problem defined by the Hamiltonian $h_B$ (Eq.(1)).
If the potentials are not regularized by form factors, the ground
state energy is not bound from below.
 
It is important to mention here that the considered two models are rather
different mainly due to the flavor-dependent factor 
$\bbox{\tau}_i\cdot\bbox{\tau}_j$ in OME model. 
This has important consequences in
predicting the baryon spectra, as discussed in Ref.\cite{gloz1}. 
It is fair to say that with suitable
adjustments and additional phenomenological parameters, both models can
reproduce the general pattern of PDG's baryon spectra.
Our objective is to find a way to distinguish them by considering
$\pi N$ and $\gamma N$ reactions. The situation is similar to
atomic and nuclear physics. Only by investigating reactions,
the dynamical content of the theoretical models can be truly tested.

With the Hamiltonian $h_B$ (Eq.(1)) defined above,
our first step 
is to solve the three-quark bound state problem
\begin{eqnarray}
  \label{three body eqn}
  h_B | B \rangle = m_B| B \rangle,
\end{eqnarray}
where $|B\rangle$ is the baryon wave function with
the label  $B$ denoting collectively the spin-parity $J^\pi$ 
and isospin $T$;
$m_B$ is the mass eigenvalue.
We use the diagonalization method developed in Ref.\cite{ogaito} to solve 
Eq.(4) by expanding
the baryon wave function  as
\begin{eqnarray}
  \label{wf expand}
  |B \rangle \rightarrow
|J^{\pi}T\rangle = \sum_{i} a^{J^\pi T}_i |J^\pi T;i \rangle,
\end{eqnarray}
where the basis states are 
appropriately antisymmetrized harmonic oscillator wavefunctions.
The coefficients $a^{J^\pi T}_i$ in Eq.(5) 
and the mass eigenvalues 
$m_{J^\pi T}$ of Eq.(4) are obtained from diagonalizing the matrix 
\begin{eqnarray}
  \label{Hij}
  H_{i,j} = \langle  J^\pi T;i | h_B | J^\pi T;j\rangle .
\end{eqnarray}
In practice diagonalization is performed within a limited number of basis states.
Then the solution of Eq.(4)
is a function of the oscillator range parameter $b$. 
We treat it as a variational parameter and find 
$b$ by  imposing the condition:
\begin{eqnarray}
  \label{variational}
  \frac{\partial m_B}{\partial b} = 0.
\end{eqnarray}
The basis state is chosen so that the mass eigenvalue  $m_{J^\pi T}$
does not change by further extension of the basis states.
In practice we include the basis states up to 11$\hbar\omega$.

The next step is to define how an external meson($M$)
can interact with the bound three-quark systems. Namely,
we need to calculate the matrix elements for
the transitions from one-baryon states($B$) 
to meson-baryon states($M'B'$)
\begin{eqnarray}
  \Gamma_{B'M',B}^\dagger({\bbox{k}}) =
  \langle  B';M' | H_{M}({\bbox{k}})  | B \rangle ,
\end{eqnarray}
where $B$ is a bound state wave function generated from the above
structure calculation, and $H_{M}({\bbox{k}})$ is an appropriate
operator describing how a meson $M$ with a momentum ${\bbox{k}}$ 
is emitted or absorbed by constituent quarks. Following the previous 
works\cite{copley,foster,koni,li,iach},
we assume that $H_M(\bbox{k})$ for $M=\pi$ is a one-body
operator which can be derived directly from 
the nonrelativistic limit of the Feynman amplitude 
$\sim \frac{f_{\pi qq}}{m_\pi}
\bar{u}_{{\bbox{p}}'}
\gamma_5\gamma^\mu k_\mu u_{{\bbox{p}}}$ for the 
$q\leftrightarrow \pi q$ transition.
To be consistent with the nonrelativistic treatment
of constituent quarks, we keep only the terms up to the order of $p/m_q$.
In coordinate space, the resulting $q + \pi \rightarrow q$ 
transition operator is
\begin{eqnarray}
H_{\pi}(\bbox{k})=\frac{i}{\sqrt{(2\pi)^32\omega_\pi}}
\frac{f_{\pi qq}}{m_\pi}\sum_{i=1}^3
e^{i{\bbox{k}}\cdot{\bbox{r}}_i} \tau^{\alpha} \bbox{\sigma}_i \cdot \left[
  \bbox{k}-\frac{\omega_\pi}{2m_q}({\bbox{p}}_i+ {\bbox{p}}'_i) \right] F(k) ,
\label{pi-quark-int}
\end{eqnarray}
where $\alpha$ denotes the $z$-component of pion isospin and
$\bbox{p}_i$ ($\bbox{p}_i'$) is  the derivative operator acting on 
the initial (final) baryon wave function; $\bbox{k}$ and
$\omega_\pi$ are the momentum and energy of pion,
respectively. 

We fix the $\pi qq$ coupling constant by assuming
that  the $\pi NN$ vertex function can be calculated from
Eqs.(8)-(9) using the $(0s)^3$ nucleon wavefunction.
We then find that $
  \frac{f_{\pi qq}}{m_\pi} = \frac{3}{5} \frac{g_{\pi NN}}{2M_N}$,
where $M_N$ is the observed mass of the nucleon and we use the empirical value
$g_{\pi NN}^2 / 4\pi = 14$. 
The same $f_{\pi qq}$ coupling constant is also used to evaluate the
$qq$ potential Eq.(3) of OME model. Accordingly,
we also introduce a form factor 
$F(k)$ in Eq.(9) to account for the effect due to the finite size of
constituent quarks.

For photon coupling, we calculate the $B \rightarrow B^\prime \gamma$
by assuming the commonly used $\gamma-qq$ vertex interaction. The
resulting form is
\begin{eqnarray}
  \Gamma_{B'\gamma,B}^\dagger({\bbox{k}}) =
  \langle  B';\gamma | H_{\gamma}({\bbox{k}})  | B \rangle ,
\end{eqnarray}
with
\begin{eqnarray}
H_{\gamma}(\bbox{k})=\frac{i}{\sqrt{(2\pi)^32\omega_\gamma}}
\sum_{i=1}^3\frac{e_i}{2m_q}
e^{i{\bbox{k}}\cdot{\bbox{r}}_i} 
\bbox{\epsilon}\cdot \left[i\bbox{\sigma}_i \times
  \bbox{k}-({\bbox{p}}_i+ {\bbox{p}}'_i) \right]. 
\end{eqnarray}
By using the $(0s)^3$ wave functions for $N$ and $\Delta$
and setting $m_q\sim 300$ MeV to evaluate Eqs.(10)-(11), 
one finds that the nucleon 
magnetic moments can be reproduced well and the
resulting helicity amplitudes for the $\Delta \rightarrow N\gamma$
transition are
$A_{3/2} \sim -160\times 10^{-3}$ (GeV)$^{-1/2}
$($A_{1/2} = A_{3/2}/\sqrt{3})$.
These values
of helicity amplitudes, which are about 40 $\%$ lower than the
values listed by PDG but are close to
various quark model predictions\cite{capstick, iach}, 
are consistent with
our interpretation in Ref.\cite{satolee1} that the constituent quark 
model predictions
can only be compared with the $bare$ values extracted from the data
by removing the on- and off-shell effects
due to non-resonant interactions 
in a dynamical approach based on the Hamiltonian formulation.

With the vertex functions
defined above, we can then define a hadronic Hamiltonian
for investigating
$\pi N$ and $\gamma N$ reactions. It has the following form
\begin{eqnarray}
  \label{dynamical-H}
  H = H_0 + H_I ,
\end{eqnarray}
where the free Hamiltonian can be written in a second-quantization form
\begin{eqnarray}
  \label{h0}
  H_0 = \sum_{B} \int d \bbox{p}\, \varepsilon_{B}(\bbox{p})
  b^\dagger_B(\bbox{p})b_B(\bbox{p})
  +
  \sum_{M} \int d \bbox{k}\, \omega_{M}(\bbox{k})
  a^\dagger_M(\bbox{k}) a_M(\bbox{k}).
\end{eqnarray}
Here, $b^\dagger_B$($b_B$) and $a^\dagger_M$($a_M$) are
the creation (annihilation) operators for the
baryons and mesons respectively, and
  $\varepsilon_B (\bbox{p}) =  (m_B^2 + \bbox{p}^2)^{1/2} ,
  \omega_M (\bbox{k})  = (m_M^2 + \bbox{k}^2)^{1/2}$ . 
It is important to note here that
the baryon mass $m_B$ in Eq.(13) is generated dynamically from 
solving the three-quark bound state problem defined by
Eqs.(4)-(7). Their values
could be significantly different from resonance positions listed by 
PDG.  We use the experimental value for the meson mass $m_M$. 

The interaction term in Eq.\ (\ref{dynamical-H}) 
is written in terms of the vertex functions defined by Eqs.(8)
and (10)
\begin{eqnarray}
  \label{int-H}
  H_I = \sum_{BB'M} \int d \bbox{p}d\bbox{p}' d\bbox{k}
      \left[ \langle  B' |H_{M}|B; M \rangle 
        b^\dagger_{B'}(\bbox{p}') b_{B}(\bbox{p}) a_M(\bbox{k})
        + \mbox{h.c.}\right].
\end{eqnarray}
The above interaction Hamiltonian is similar to that of 
the dynamical model developed in Ref.\cite{satolee1}, except that
the $B' \rightarrow BM$ transition amplitudes are now determined by
the predicted quark wavefunctions.
As discussed in Ref.\cite{satolee1}, 
it is a non-trivial many-body problem 
to calculate $\pi N$ and $\gamma N$ reactions with the use of $H_I$.
To obtain a manageable reaction 
theory, we follow Refs.\cite{satolee1,Kobasaoh} and apply the unitary
transformation up to the
second order in $H_I$ to derive an effective Hamiltonian. 
The essence of the unitary transformation method 
applied in Ref.\cite{satolee1}
is to absorb the unphysical transition $B \rightarrow M' B'$ with 
$m_{B} < m_{B'}+ m_{M'}$ into non-resonant potentials.
The resulting effective Hamiltonian then takes the following form
\begin{eqnarray}
  \label{effhi}
  H_{\rm eff} &=& H_0+ \Gamma + \Gamma^{\dagger} + v , 
\end{eqnarray}
where $H_0$ is defined in Eq.\ (\ref{h0}). The vertex $\Gamma^\dagger$ 
contains only the physical decay
process $B \rightarrow M'B'$ with $m_{B} > m_{M'} + m_{B'}$
\begin{eqnarray}
  \Gamma^\dagger &=& \sum_{MBB^\prime}
  \int d \bbox{k} d \bbox{p} d \bbox{p}^\prime\ 
  \langle B';M' | H_{Mqq} |B \rangle
  b^\dagger_{B'}(\bbox{p}') a^\dagger_{M'}(\bbox{k}') b_{B}(\bbox{p})
\nonumber \\
 &\times&  \theta(m_{B} - (m_{B'} + m_{M'})),
\end{eqnarray}
where $\theta(x) = 1(0)$ for $x > 0 ( x< 0)$.
The non-resonant $MB \rightarrow M'B'$ two-body interactions are defined by
\begin{eqnarray} 
  v = \sum_{MM^\prime BB^\prime} \int d \bbox{k} d \bbox{k}^\prime
 d \bbox{p} d \bbox{p}^\prime  
   \langle B^\prime; M^\prime  | \hat{v} |  B;M \rangle
  a^{\dagger}_{M'}(\bbox{k}^\prime) a_M(\bbox{k})
  b^{\dagger}_{B'}(\bbox{p}^\prime) b_B(\bbox{p}) .
\end{eqnarray} 
The construction of the matrix elements of the nonresonant potential
$\hat{v}$ is explained in Ref.\cite{satolee1}. Typically, it consists of
the energy-forbidden $s$-channel terms such as $\pi N
\rightarrow N \rightarrow \pi N$ and $\pi \Delta 
\rightarrow \Delta \rightarrow \pi \Delta$,
and particle-exchange terms such as the usual crossed nucleon and
$\Delta$ terms and the $\rho$- and $\omega$-exchange terms.  

By using the standard projection operator method\cite{satolee1},
it is straightforward to derive 
from the effective Hamiltonian Eq.\ (\ref{effhi})
a calculational framework for $\pi N$ and $\gamma N$ reactions
leading to various final meson-nucleon final states.
The resulting transition operator can be written as
\begin{eqnarray}
  \label{tmat}
  T_{\alpha,\beta} = t_{\alpha,\beta} +
  \sum_{i,j}\tilde{\Gamma}^\dagger_{\alpha,N^*_i}\left
  [ D^{-1}(E)\right]_{i,j}\tilde{\Gamma}_{N^*_j,\beta} .
\end{eqnarray}
Here $\alpha, \beta$ denote the meson-baryon states 
such as $\gamma N$, $\pi N, \eta N$
and $ \pi \Delta$.
$N^*_i$ are mass eigenstates of Eq.\ (\ref{three body eqn}).
The first term in Eq.\ (\ref{tmat}) is the non-resonant amplitude 
determined only by 
the non-resonant interaction $v$
\begin{eqnarray}
  \label{bg-tmat}
  t_{\alpha,\beta}= v_{\alpha,\beta} + \sum_{\gamma}
  v_{\alpha,\gamma}G^0_{\gamma}(E) t_{\gamma\beta},
\end{eqnarray}
with
\begin{eqnarray}
  \label{bg-propag}
  \left[ G^0_\gamma (E)\right]^{-1} =
  E - \varepsilon_{B_\gamma}(\bbox{p}) - \omega_{M_\gamma}(\bbox{k})
  + i \epsilon .
\end{eqnarray}
The second term in Eq.\ (\ref{tmat}) is the resonant term determined by the dressed $N^*$
propagator and the dressed vertex functions:
\begin{eqnarray}
  \label{nsprop}
  \left[D(E)\right]_{i,j} &=&(E - m_{N^*_i})\delta_{ij} - \Sigma_{i,j}(E), \\
  \tilde{\Gamma}_{N^*_i,\alpha} &=& \sum_{\gamma}\Gamma_{N^*_i,\gamma}
  \left[\delta_{\gamma\alpha} +
    G^0_{\gamma}(E)t_{\gamma,\alpha}\right] , \\
  \tilde{\Gamma}^\dagger_{\alpha,N^*_i} &=&
  \sum_{\gamma}\left[\delta_{\gamma\alpha}
    +t_{\alpha,\gamma}G^0_{\gamma}(E)
  \right]\Gamma^\dagger_{\gamma,N^*_i}.
 \end{eqnarray}
In Eq.\ (\ref{nsprop}), the $N^*$ self-energy is defined by
\begin{eqnarray}
  \label{nsSelf}
  \Sigma_{i,j}(E) = \sum_{\gamma} \Gamma_{N^*_i,\gamma}
  G_{\gamma}^0(E)\tilde{\Gamma}^\dagger_{\gamma, N^*_j} .
\end{eqnarray}

We now note that
the above coupled equations relate the
full scattering amplitude $T_{\alpha,\beta}$ Eq.(18)
nonperturbatively to the quark wavefunctions calculated from
the considered constituent quark model.
In particular, the decay widths, as listed by PDG,
correspond to the dressed vertex $\tilde{\Gamma}_{N^*,\gamma}$ defined
by Eq.(22), and hence they are expected to be different dynamically
from the  bare vertex functions $\Gamma_{N^*,\gamma}$ 
which are calculated from using Eqs.(8)-(11).
It is also clear that the accuracy of the
non-resonant interaction $v_{\alpha,\beta}$, which determine the
nonresonant amplitude $t_{\alpha,\beta}$ via Eq.(19),
plays an important role in identifying the resonant amplitude of
Eq.(18) from the data.
It is therefore also essential to have it calculated from the
same quark model. This can be achieved in our approach through the
use of unitary transformation method, as discussed above(the
developed method can be
straightforwardly extended to account for the interactions
due to heavy mesons).
Without such a consistent treatment of both the resonant and 
non-resonant interactions, it is difficult to draw conclusions.

\begin{figure}
  \begin{center}
    \begin{tabular}{cc}
      \epsfig{file=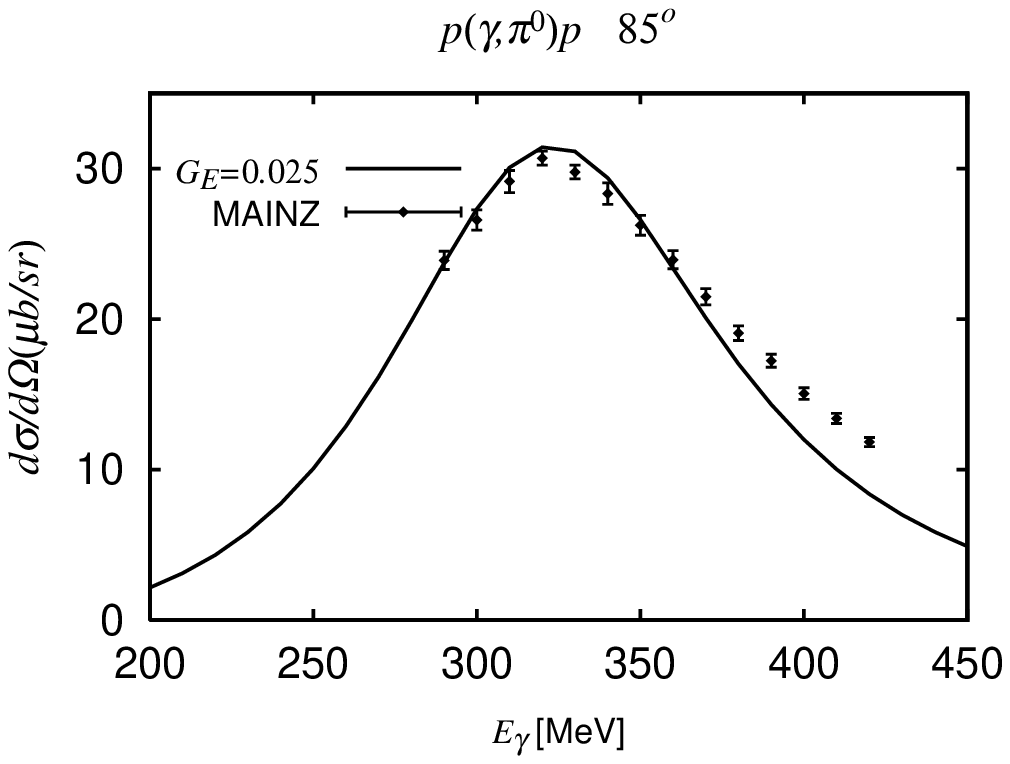,width=5cm} &
      \epsfig{file=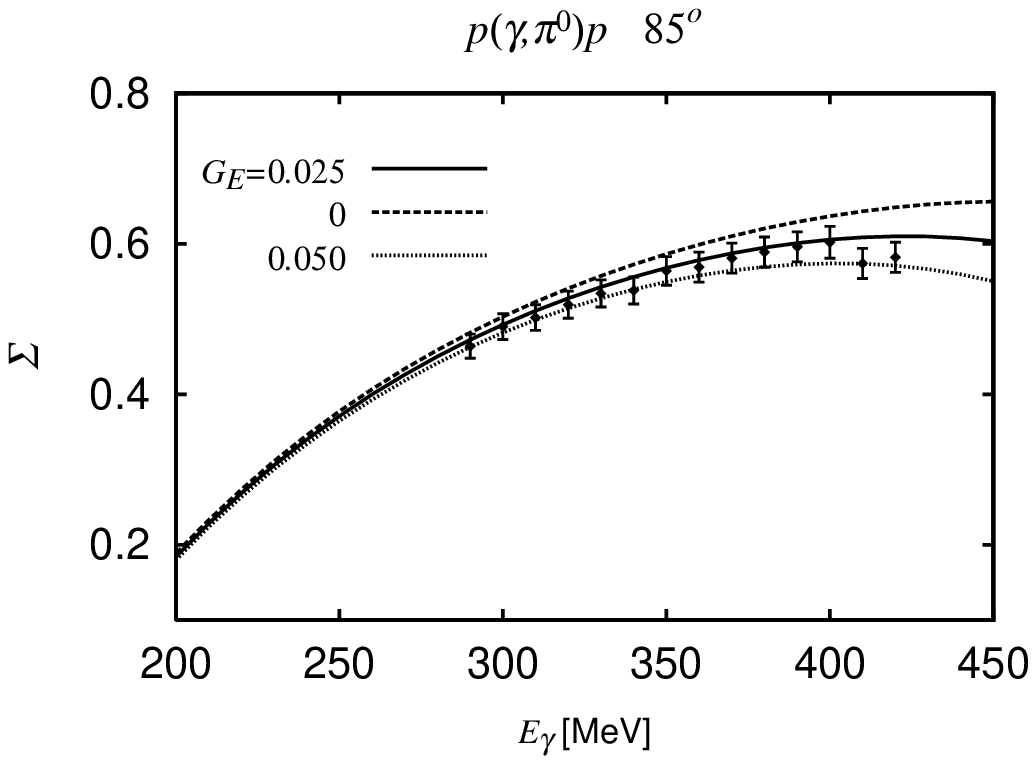,width=5cm} \vspace{1mm}\\
      (a) & (b)
    \end{tabular}
  \end{center}
  \caption[]{The recent Mainz data[19] of 
differential cross section and
photon asymmetry $\Sigma$ of the $p(\gamma,\pi^0)p$ reaction are
compared with the predictions from the SL model.
$G_E$ is the strength of the $E_{1^+}$
amplitude of the $\Delta \rightarrow \gamma N$ transition. }
\end{figure}
\begin{figure}
  \centerline{\epsfig{file=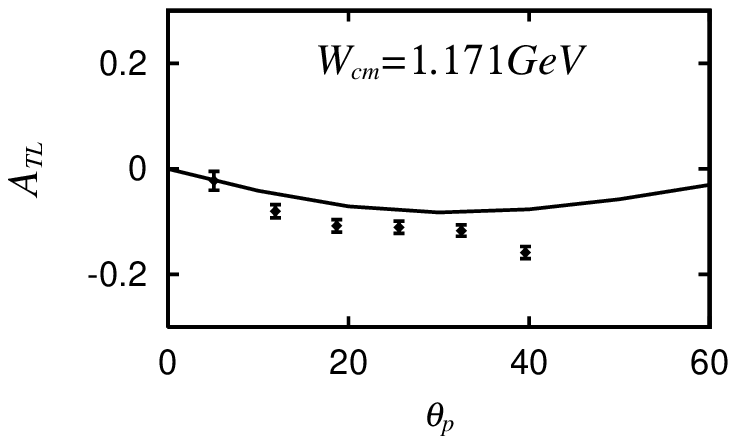,width=6cm} 
 \epsfig{file=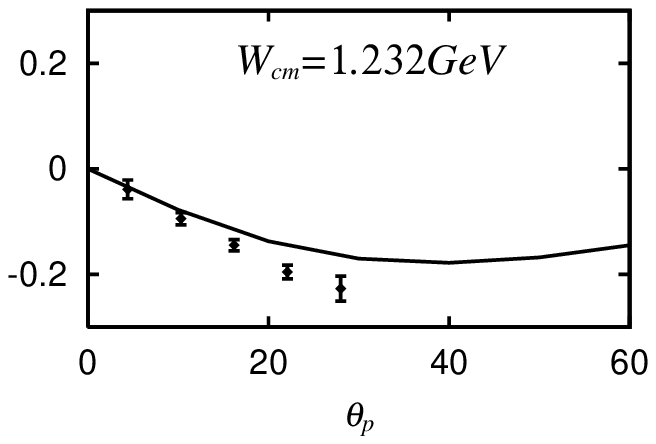,width=6cm}}
  \caption[]{The MIT-Bates data[20] of asymmetry $T_{TL}$
of the $p(e,e'\pi^0)$ reaction at $Q^2=0.126$ (GeV/c)$^2$ are
compared with the predictions by the SL model.}
\end{figure}
We now turn to discussing our findings. First we consider the
$\Delta$ state in $P_{33}$ channel. For this investigation,
we only have one $N^*(=\Delta)$ state
and two channels $\alpha,\beta = \pi N, \gamma N$. 
Eqs.(18)-(24) are then reduced to the
set of equations developed in 
Ref.\cite{satolee1}(called SL model from nowon).
The accuracy of the SL model are illustrated in Figs.1-2.
We see that the predictions are in good agreement with 
the recent data from Mainz\cite{mainz} and MIT-Bates\cite{mit}.
It is therefore sufficient to adjust
the parameters characterizing the 
the $qq$ potentials Eqs.(3) or (4) and the
constituent quark form factor $F(k)$ for
the vertex interaction Eq.(9) to fit the bare $\Delta$ parameters 
of the SL model. This is a rather restricted procedure since there are
only about 4 parameters for each of the considered OGE and OME models.

In Fig.3, we show that the $\Delta \rightarrow \pi N$ form factor
calculated from using Eqs.(8)-(9) agree
to a large extent with 
the phenomenological bare form factor(dashed curve) of SL model.
The predicted mass(via Eq.(4)) is also required to
be the bare mass $m_\Delta=1300$ MeV of SL model.
\begin{figure}
      \centerline{\epsfig{file=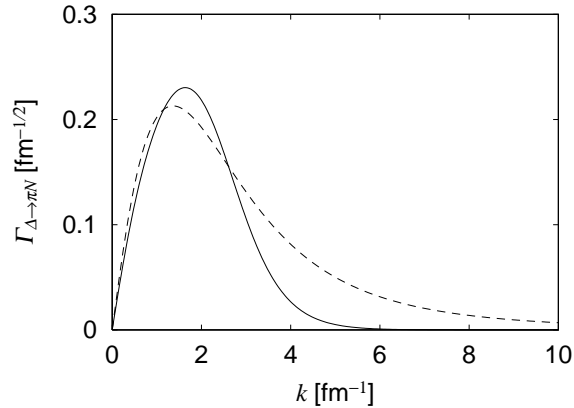,width=8cm}} 
  \caption[]{The $\Delta \rightarrow \pi N$ vertex 
function(solid curve) determined within the OGE model
is compared with
     the phenomenological form(dashed curve) of
    the SL model[1].} 
  \label{fig:piND-vertex}
\end{figure}

Once the parameters of the considered constituent quark model
are determined,  we then can generate the mass $m_B$ and
quark wavefunctions for all possible partial waves by solving
Eqs.(4)-(7). The relevant vertex
functions can then be calculated by using Eqs.(8)-(11) and
the $\pi N$ and $\gamma N$ reaction amplitudes 
can be predicted by solving Eqs.(18)-(24).
The comparison with the
data will tell us whether the considered constituent quark model
is correct. We have explored this by investigating
the $\pi N$ scattering in $S_{11}$ partial wave.
The calculation was done by considering
two $N^*$ and three channels $\pi N, \eta N$ and $\pi \Delta$. 
The predicted $\pi N$ phase shifts are shown in
Fig.4a. We see that neither OGE(solid curve)
 or OME(dashed curve) models can describe the data(open circles) 
in the entire energy region. 

As an attempt to  improve the fit to the $S_{11}$ amplitude, 
we have also explored the mixture of OGE and OME models.
It turns out that such a hybrid model also fails, 
mainly due to the very 
disruptive tensor component of the OME model in determining the
phases of wave functions.

We therefore turn to 
investigating a purely phenomenological model. 
By analyzing an analytical model\cite{yoshimoto},
we have found that the data of $\Delta$ excitation
and $S_{11}$ $\pi N$ scattering seem to 
favor a tensor term due to one-gluon-exchange
and a spin-spin interaction due to one-meson-exchange.
This has guided us to explore many phenomenological
models. For example, we have found that the $\pi N$ $S_{11}$ amplitudes
can be much better described
by the following phenomenological model
\begin{eqnarray}
V^{phe}_{qq}(q) = \bbox{\sigma}_1\cdot\bbox{\sigma}_2
\bbox{\tau}_1\cdot\bbox{\tau}_2 V_{\sigma\tau}(q)
+S_{12}V_{T}(q)
\end{eqnarray}
The results are shown in Fig.4b. We see that the general feature
of the data can now be reproduced. The remaining discrepancies
perhaps can be removed if we refine various form factors in
the $qq$ potential and $N^*\rightarrow \pi N,\eta N,\pi\Delta$
vertex functions(Eq.(9)).
\begin{figure}
  \begin{center}
    \begin{tabular}{cc}
      \epsfig{file=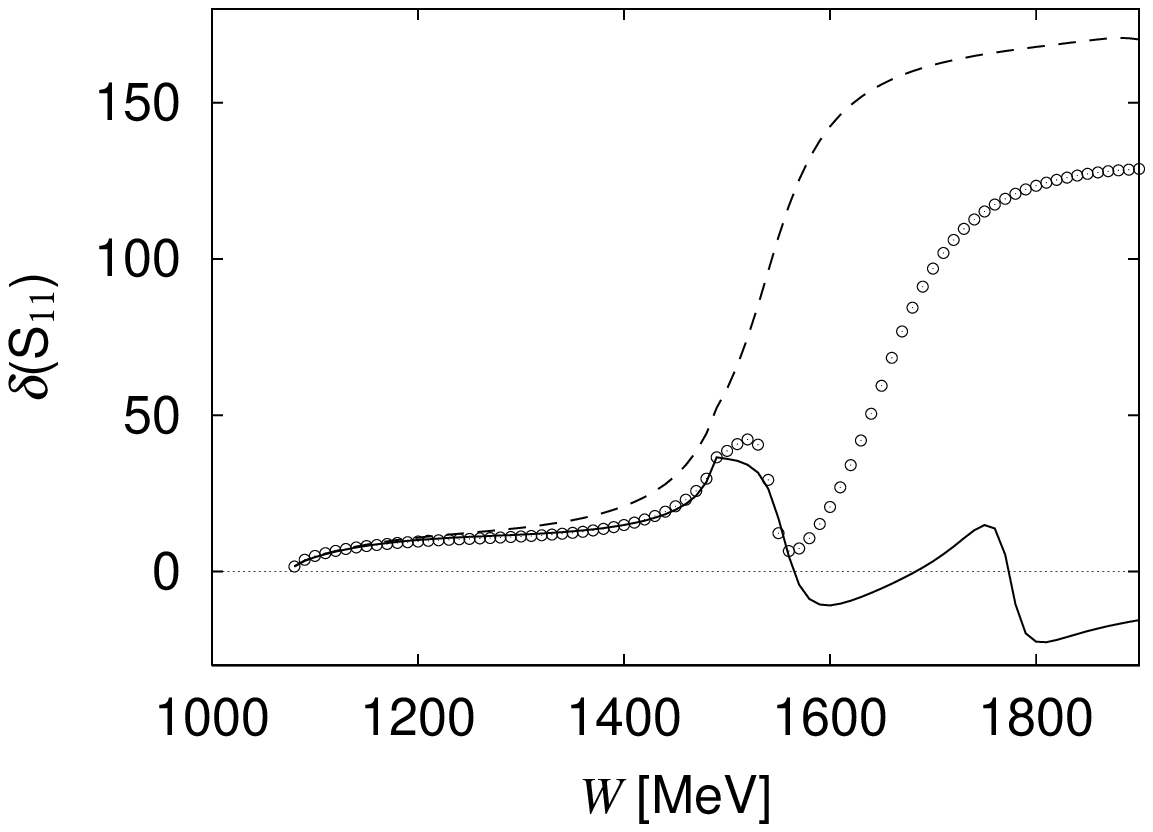,width=5cm} &
      \epsfig{file=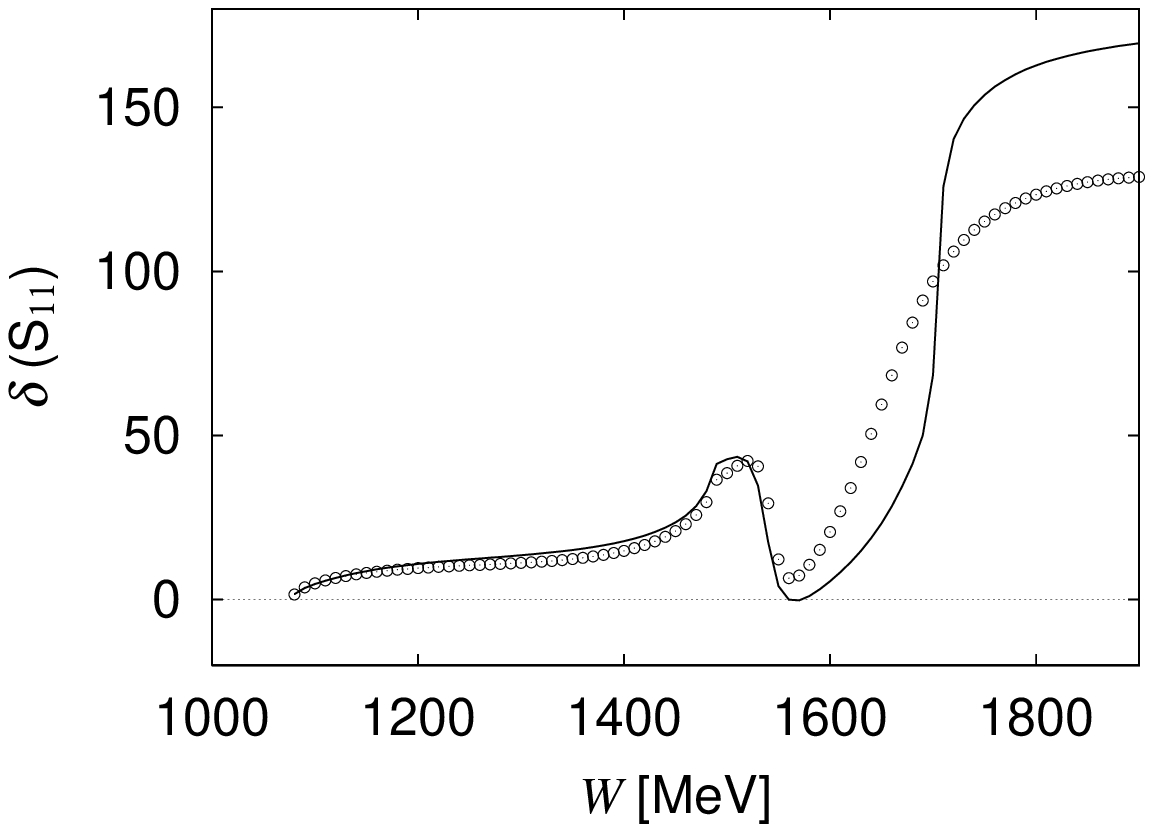,width=5cm} \vspace{1mm}\\
      (a) & (b)
    \end{tabular}
  \end{center}
  \caption[]{ The predicted $\pi N$ phase shifts are compared with the
data(open circles):(a) from
OGE(solid curve) and OME(dashed curve) models, (b) from phenomenological
model Eq.(25).}
    \label{fig:phe-amp}
\end{figure}

In conclusion, we have developed a Hamiltonian formulation
for relating the dynamics of $\pi N$ and $\gamma N$ reactions to
the constituent quark models. The approch has been applied to
investigate the $\Delta$ and $N^*(S_{11})$ resonances.
It is found that while the $\Delta$ excitation can be described
satisfactory, 
the $\pi N$ scattering in $S_{11}$ channel can not be
described by the constituent quark models based on either the 
one-gluon-exchange(OGE) or one-meson-exchange(OME) mechanisms.
The data seem to favor the
spin-spin interaction due to one-meson-exchange and the tensor
interaction due to one-gluon-exchange. A phenomenological
quark-quark potential has been constructed to reproduce the
$S_{11}$ amplitude. 

To end, we would like to point out that the multi-channels
multi-resonances parameterization of $\pi N$ reaction amplitudes,
first developed by Cutkosky and his collaborators\cite{cut}
and has been revived recently\cite{var}, can be
derived\cite{lee} from Eqs.(18)-(24).
The bare parameters associated with this phenomenological 
model can have dynamical interpretations
within our Hamiltonian formulation. Consequently, it may be
more advantageous to apply this parameterization to analyze the
forthcoming data. The extracted bare $N^*$ parameters can be
used as the data for exploring the constituent quark interactions 
using the procedures detailed in Ref.\cite{yoshimoto} and 
outlined in this contribution.

\vspace*{-2pt}

\section*{Acknowledgments}
This work was supported 
U.S. DOE Nuclear Physics Division Contract No. W-31-109-ENG-38
and JSPS Grant-in-Aid for Scientific Research (C) 12640273.


                                %
                                %
                                %


\vspace*{-9pt}

\section*{References}
\begin{thebibliography}{99}
\bibitem{satolee1}
T. Sato and T.-S. H. Lee, Phys. Rev. {\bf C54}, 2660 (1996).
\bibitem{satolee2}
T. Sato and T.-S. H. Lee, contribution to 16th International Conference
, March 6-10, 2000, and to be published.
\bibitem{yoshimoto}
T. Yoshimoto, T. Sato, M. Arima, and T.-S. H. Lee, Phys. Rev. C,
June issue, 2000.

\bibitem{copley}
L. A. Copley, G. Karl, and E. Obryk, Nucl. Phys. {\bf B13}, 303 (1969).
\bibitem{foster}
F. Foster and G. Hughes, Z. Phys. {\bf C14}, 123 (1982).
\bibitem{koni}
R. Koniuk and N. Isgur, Phys. Rev. Lett. {\bf 44}, 845 (1980);
Phys. Rev. {\bf D21}, 1868 (1980).
\bibitem{li}
Zhenping Li and F. E. Close, Phys. Rev. {\bf D42}, 2207 (1990).
\bibitem{iach}
R. Bijker, F. Iachello, and A. Leviatan, Ann. Phys.(N.Y.) 
\bibitem{caps2}
S. Capstick and W. Roberts, Phys. Rev. {\bf D47}, 1994 (1993); 
{\it ibid} {\bf D49}, 4570 (1994);{\it ibid} {\bf D57}, 4301 (1998).
\bibitem{isgur} 
N. Isgur and G. Karl, Phys. Rev. {\bf D18}, 4187 (1978);
{\it ibid} {\bf D19}, 2653 (1979).
\bibitem{caps1}
S. Capstick and N. Isgur, Phys. Rev. {\bf D34}, 2809 (1986).
\bibitem{capstick}
S. Capstick, Phys. Rev. {\bf D46}, 1965 (1992);
{\bf 46}, 2864 (1992)
\bibitem{gloz1}
L. Ya. Glozman and D. O. Riska, Phys. Rep. {\bf 268}, 263 (1996).
\bibitem{gloz2}
L. Ya. Glozman, Z. Papp, W. Plessas, K. Varga, and R.F. Wagenbrunn,
Nucl. Phys. {\bf A623}, 90c (1997).
\bibitem{ogaito}
T. Ogaito, T. Sato, and M. Arima, unpublished.
\bibitem{Kobasaoh}
M. Kobayashi, T. Sato and H. Ohtsubo, Prog. Theor. Phys. {\bf 98}, 927 (1997).
\bibitem{cut}
R.E. Cutkosky, C.P. Forsyth, R.E. Hendrick, R.L. Kelly, Phys. Rev.
{\bf D20}, 2839 (1979)

\bibitem{var}
T.P. Vrana, S.A. Dytman, and T.-S. H. Lee, Phys. Rep. Vol.328, 181 (2000)

\bibitem{mainz}
R. Beck et al, Phys. Rev. {\bf C61},035204-1 (2000)

\bibitem{mit}
C. E.  Vellidis et al, page 105, Proceedings of the
second workshop on Electronuclear Physics and the BLAST Detector,
editted by R. Alarcon and R. Milner, World Scientific(1999)

\bibitem{lee}
T.-S. H. Lee, unpublished, (1996).
\end{thebibliography}
\end{document}